# Impact of Fe$_{80}$B$_{20}$ insertion on the properties of dual-MgO perpendicular magnetic tunnel junctions

Enlong Liu[1], Taeyoung Lee[2] and Hyunsoo Yang[1]

[1] Department of Electrical and computer Engineering, National University of Singapore, 117576, Singapore
[2] GLOBALFOUNDRIES Singapore Pte. Ltd., Singapore 738406, Singapore
E-mail: eleyang@nus.edu.sg

**Abstract**

We explore the impact of Fe$_{80}$B$_{20}$ inserted at both Co$_{20}$Fe$_{60}$B$_{20}$/MgO interfaces of dual-MgO free layers (FLs) in bottom-pinned magnetic tunnel junctions (MTJs). MTJ stacks are annealed for 30 min at 350 °C and 400 °C in a vacuum after film deposition. Current-in-plane tunneling measurements are carried out to characterize magnetotransport properties of the MTJs. Conventional magnetometry measurements and ferromagnetic resonance are conducted to estimate the saturation magnetization, the effective perpendicular anisotropy field and the Gilbert damping of dual-MgO FLs as a function of the Fe$_{80}$B$_{20}$ thickness and annealing temperatures. With ultrathin Fe$_{80}$B$_{20}$ (0.2 – 0.4 nm) inserted, perpendicular magnetic anisotropy (PMA) of FLs increases with similar tunnel magneto-resistance (TMR) and low damping values. As Fe$_{80}$B$_{20}$ layer thickness further increases (0.6 – 1.2 nm), both TMR and PMA degrade, and damping increases dramatically. This study demonstrates a novel approach to tune properties of MTJ stacks with dual-MgO FLs up to 400 °C annealing, which enables MTJ stacks for various applications.

## 1. Introduction

Magnetic tunnel junctions (MTJs) with perpendicular magnetic anisotropy (PMA) have been studied in the recent decades as the crucial element for next generation memory applications, such as spin-transfer-torque and spin-orbit-torque magnetic random access memory (STT/SOT-MRAM), due to their non-volatility, energy effectiveness, high endurance, and scalability [1–3]. Many efforts have been made in the engineering of the data-storage layer in MTJs, i.e. the free layer (FL), whose magnetic moment can be switched by the writing current. Especially, dual-MgO FLs with a structure as MgO/CoFeB/spacer/CoFeB/MgO have been under intense development [4–6]. On the one hand, dual-MgO FLs can provide high PMA and low damping to guarantee the high thermal stability and low switching current, after scaling-down of MTJ devices [7]. On the other hand, the aformentioned properties of dual-MgO FLs can be maintained after post-annealing up to 400 °C, which is required for the CMOS back-end-of-line (BEOL) process [8].

To further optimize dual-MgO FLs performance, previous studies focused on different topics. Among them, non-magnetic spacer engineering and element composition effects have drawn lots of interest. Researches on non-magnetic spacer sandwiched between two CoFeB layers in dual-MgO FLs have been widely conducted. Materials such as Mo [9–11], Ta [8,12], and W [13–15] were explored as the spacer to identify its impact on PMA and damping before and after annealing. Other works examined the effect of element (Fe or B) composition in CoFeB layers in MTJ stacks on several parameters, including tunnel magneto-resistance (TMR) [16], PMA [17–19], and annealing stability [20]. It has been demonstrated that a thickness gradient in the B content can modify the properties of CoFeB/MgO bilayer system such as damping and anisotropy [21]. PMA of the FL was also reported to be improved with increasing the Fe composition in CoFeB, on which its TMR is almost independent [22]. However, it is still an open question how a gradient of Fe in dual-MgO FLs impacts the overall properties of MTJ stacks. In such a case, the PMA of dual-MgO FLs would benefit from an increased Fe concentration, while the TMR of the MTJs is expected to be improved due to the formation of Fe/MgO/Fe interface after annealing [23].

Here we propose an insertion of ultrathin $Fe_{80}B_{20}$ (hereafter FeB) at the interface of $Co_{20}Fe_{60}B_{20}$/MgO in dual-MgO FLs to achieve tunable magnetic properties and annealing stability. By optimizing the thickness of the FeB insertion layers at both CoFeB/MgO interfaces, a large PMA can be obtained after 350 °C annealing and further improved at 400 °C, which is accompanied with a low damping constant. The result of a low saturation magnetization, large anisotropy field and low damping at the same time in the FeB-inserted dual-MgO FLs makes it promising for low switching currents in MTJ devices without reducing the thermal stability [24]. In addition, the tunability of FL performance by FeB insertion enlarges its potential for various spintronic applications where CoFeB-based MTJs are present, such as SOT-MRAM, spin logic devices and STT nano-oscillators.

## 2. Experimental

Bottom-pinned perpendicular MTJs with [Co/Pt] multilayers as a perpendicular synthetic antiferromagnet (p-SAF) were *in-situ* deposited at room temperature by magnetron sputtering on W/Ru/W/Ru/W bottom electrodes (BE) and capped by the Ta/Ru top electrode (TE) in a ULVAC Magest S200 multi-chamber machine. All samples were first annealed with a 0.5 T magnetic field perpendicular to film plane in a magnetic vacuum annealing oven at 350 °C for 30 min. The TMR and resistance-area product (RA) of the MTJ stack was measured via current-in-plane tunnelling method (CIPT) [25]. The hysteresis loops of blanket stacks were measured by a vibrating sample magnetometer (VSM) with the magnetic field perpendicular to the sample plane. Field-modulated ferromagnetic resonance (FMR) measurements with the frequency range of 10-25 GHz were conducted to extract the resonance field and linewidth of the FL versus the frequency, from which the effective perpendicular anisotropy field and Gilbert damping of the FL can be estimated. All FMR measurements were conducted with samples placed film-side down on a coplanar waveguide in an electromagnet with a field range up to 0.5 T and perpendicular to the sample plane. The same batch of samples were then annealed at 400 °C for 30 min to study the annealing impact.

## 3. Results and discussion

### 3.1 Stack characterization without FeB insertion

The detailed stack structure of MTJs used in this study is provided schematically in figure 1(a). It consists of (thickness in nm):
- Hard layer (HL): Pt (5)/[Co (0.25)/Pt (0.2)]/Co (0.6)
- Reference layer (RL): Co (0.6)/Pt (0.2)/Co (0.3)/Pt (0.2)/Co (0.5)/W (0.3)/$Co_{20}Fe_{60}B_{20}$ (0.8)

- Free layer (FL): $Co_{20}Fe_{60}B_{20}$ (1.2)/W (0.4)/ $Co_{20}Fe_{60}B_{20}$ (0.8).

The FL is sandwiched between the MgO tunnel barrier and the 2nd MgO layer to form a dual-MgO structure. The magnetic hysteresis loop of the full stack in figure 1(b) indicates good PMA in each functional layer after 350 °C annealing. From the minor hysteresis loop shown in figure 1(c), PMA of the FL is still maintained after 400 °C annealing, and the coercive field also increases slightly. The reduction of the saturation magnetization per area ($M_s \cdot t$) of the FL after 400 °C annealing can be attributed to magnetic dead layer formation, which will be discussed in the following sections.

After 400 °C annealing, a sloped plateau is observed around 500 mT in the red curve in figure 1(b), indicating a decreased PMA in the RL [26]. In addition, the TMR of the stack is reduced from 110% to 50%. Since the FL PMA is maintained, the TMR reduction is attributed to a PMA loss in the RL. Thus, the impact of FeB insertion in dual-MgO FLs on TMR is studied in stacks after 350 °C annealing.

*3.2 Impact of FeB insertion on TMR and RA*

Figure 2(a) and (b) show schematically the FeB insertion position in the dual-MgO FL. Top FeB is inserted between the 2nd MgO layer and CoFeB above W spacer, while bottom FeB is between the MgO tunnel barrier and CoFeB below W spacer. To eliminate the difference in the thickness of FL after insertion, the total thickness of FeB insertion plus remaining CoFeB is kept at 0.8 nm and 1.2 nm for layers above and below W spacer, respectively. As such, the thickness of the top FeB insertion layer is chosen as 0.2, 0.4, 0.6 and 0.8 nm. For the bottom FeB insertion layer, its thickness options are 0.2, 0.4, 0.6, 0.8, 1.0 and 1.2 nm.

The TMR and RA values as a function of the FeB insertion layer at different CoFeB/MgO interfaces are summarized in figure 2(c) and (d), respectively. The trend for both TMR and RA is similar regardless of FeB insertion position. The TMR value is similar for the thickness of FeB < 0.4 nm, but the RA value reduces for thin FeB insertion < 0.6 nm and increases with thicker FeB. In addition, it is found that top FeB insertion leads to a more pronounced RA reduction. This phenomenon indicates that with top FeB insertion, the contribution to RA from the 2nd MgO layer can be reduced. Since the MgO tunnel barrier (0.85 nm) is thicker than the 2nd MgO layer (0.7 nm), its RA change due to bottom FeB insertion is not as significant as in the top FeB case. Overall, a similar TMR value at low RA is realized when ultrathin FeB (0.2 – 0.4 nm) is inserted, suggesting that the formation of Fe/MgO interface benefits magnetotransport properties of MTJ stacks.

The TMR starts to drop when FeB thicker than 0.6 nm was inserted at the bottom CoFeB/MgO interface, but the RA value does not show any significant change. This TMR drop is attributed mainly to a lower spin polarization when thicker FeB replaces CoFeB in the FL [27].

*3.3 Impact of FeB insertion and annealing on FL magnetic properties*

In figure 1(c), an example of hysteresis loop of the FL is shown. The saturation magnetization per area of the FL can be estimated by using $M_s \cdot t = m/A$, where $m$ is the magnetic moment, $t$ is the thickness of FL, and $A$ is the sample area. The effective anisotropy field ($\mu_0 H_k^{eff}$) is derived from FMR measurements, as shown by exemplary data in figure 3(a) from dual-MgO FLs without FeB insertion, with 0.2 nm top FeB insertion and 0.2 nm bottom FeB insertion, all after 350 °C annealing. For each sample, the power absorption by FL versus the applied field scan is measured at various frequencies, from which the ferromagnetic resonance field ($\mu_0 H_{res}$) and linewidth ($\mu_0 \Delta H$) is estimated. The relation between $\mu_0 H_{res}$ and frequency $f$ is described by the Kittel equation for the out-of-plane applied field [28]:

$$f = \frac{\gamma}{2\pi}\left(\mu_0 H_{res} + \mu_0 H_k^{eff}\right) \tag{1}$$

where $\gamma$ is the gyromagnetic ratio. From figure 3(a), the x-intercept is the $\mu_0 H_k^{eff}$. From $M_s \cdot t$ and $\mu_0 H_k^{eff}$, the effective perpendicular anisotropy energy can be calculated as

$$K_U^{eff} \cdot t = \frac{1}{2}\mu_0 H_k^{eff}(M_s \cdot t). \tag{2}$$

Figure 4 summarized the results calculated by the above method. First, $M_s \cdot t$ of FLs can be described simply by $M_s \cdot t = M_s^{CoFeB} \cdot 2.0 - t_{FeB} \cdot (M_s^{CoFeB} - M_s^{FeB})$, where $t_{FeB}$ is the inserted thickness of FeB. It is found in figure 4(a) and (b) that $M_s \cdot t$ decreases with thicker inserted FeB, i.e. the slope is negative. Thus, in our stack $M_s^{FeB}$ is smaller than $M_s^{CoFeB}$. Next, for the top FeB insertion case in figure 4(a), the amount of change in $M_s \cdot t$ at the same FeB insertion thickness after 350 °C annealing is larger than that of the bottom insertion case in figure 4(b). This indicates that the top FeB inserted layer is more damaged than the bottom inserted FeB. Another major difference between top and bottom insertion cases is a larger $M_s \cdot t$ reduction after 400 ºC annealing compared to that after 350 °C annealing, with thick bottom FeB inserted (> 0.4 nm). As the bottom-inserted FeB is less damaged, it still reduces the $M_s \cdot t$ value on further annealing at 400 ºC. However, for the top inserted case, there is not much change after 400 ºC annealing, supporting that the top FeB insertion layer is damaged, mainly due to the 2nd MgO deposition.

The FeB thickness dependence of $\mu_0 H_k^{eff}$ and $K_U^{eff} \cdot t$ can be discussed together. In the top FeB insertion case after 350 ºC annealing, $\mu_0 H_k^{eff}$ in figure 4(c) increases monotonically with FeB insertion. Due to the reduction in $M_s \cdot t$, however, this increase cannot lead to a higher PMA. In figure 4(e), the PMA of FL after 350 ºC  annealing changes little, and even slightly decreases with thicker FeB. However, after 400 ºC annealing, $\mu_0 H_k^{eff}$ overall increases without any clear dependence on the FeB thickness, and the PMA of FL reaches the maximum value when 0.2 nm FeB is inserted.

On the other hand, the behavior of $\mu_0 H_k^{eff}$ and $K_U^{eff} \cdot t$ in bottom-inserted FeB cases is different. In general, the PMA can be improved more significantly with bottom-inserted FeB than top-inserted FeB cases. In 350 ºC  annealing cases, both $\mu_0 H_k^{eff}$ and $K_U^{eff} \cdot t$ increase till 1.0 nm FeB insertion as shown in figure 4(d) and figure 4(f), respectively. While in 400 ºC annealing cases, $K_U^{eff} \cdot t$ is significantly improved in thin FeB (0.2 – 0.6 nm) cases. To summarize, the impact of FeB on the PMA of FL differs according to the insertion position. Regardless of annealing temperature, FeB insertion at the bottom CoFeB/MgO interface induces a larger PMA, probably due to less damage and the formation of Fe/MgO interface. By changing the FeB insertion position and thickness, magnetic properties of dual-MgO FL can be tuned in a wide range.

It can be noticed that the vertical error bar of anisotropy field is huge when the FeB thickness is 0.8 nm in the top insertion case, and 1.0 or 1.2 nm in bottom insertion cases. It reflects that the resonance field is difficult to be determined precisely in those cases, and thus the fits of Eq.(1) contains large uncertainties. It is probably due to a large damping constant in thick FeB insertion cases, which will be discussed in the following section.

*3.4 Impact of FeB insertion layer on FL damping*

**Table 1.** Comparison of magnetotransport and magnetic properties of dual-MgO FLs with different FeB insertion after 350 °C 30 min annealing.

| FL types | TMR % | RA Ω·μm$^2$ | $M_s \cdot t$ ×10$^{-5}$ A | $\mu_0 H_k^{eff}$ mT | $K_U^{eff} \cdot t$ mJ·m$^2$ | $\alpha$ ×10$^{-3}$ |
|---|---|---|---|---|---|---|
| MgO/CoFeB(1.2)/W/CoFeB(0.8)/MgO | 109.3 | 8.5 | 184.0 ± 2.7 | 320 ± 1 | 0.294 ± 0.004 | 12.5 ± 2.6 |
| MgO/CoFeB(1.2)/W/CoFeB(0.6)/**FeB(0.2)**/MgO | 114.9 | 6.9 | 177.7 ± 3.1 | 340 ± 5 | 0.302 ± 0.007 | 15.0 ± 4.9 |
| MgO/**FeB(0.2)**/CoFeB(1.0)/W/CoFeB(0.8)/MgO | 108.6 | 8.3 | 181.2 ± 1.7 | 349 ± 2 | 0.316 ± 0.006 | 11.3 ± 5.0 |
| MgO/**FeB(0.2)**/CoFeB(1.0)/W/CoFeB(0.6)/**FeB(0.2)**/MgO | 108.2 | 7.5 | 175.7 ± 2.6 | 381 ± 2 | 0.335 ± 0.005 | 12.1 ± 0.7 |
| MgO/**FeB(0.4)**/CoFeB(0.8)/W/CoFeB(0.6)/**FeB(0.2)**/MgO | 110.5 | 8.6 | 168.3 ± 2.9 | 434 ± 2 | 0.365 ± 0.006 | 4.4 ± 0.7 |

In order to evaluate the effect of FeB insertion on the FL damping, FMR measurements are conducted. In figure 3(b), $\mu_0 \Delta H$ versus the external excitation frequency for the three samples was plotted. The linewidth of the resonance is linear in frequency:

$$\mu_0 \Delta H = \mu_0 \Delta H_0 + \frac{4\pi\alpha}{\gamma} f \qquad (3)$$

where $\mu_0 \Delta H_0$ is the inhomogeneous linewidth broadening and $\alpha$ is the Gilbert damping coefficient.

Figure 5 summarizes $\alpha$ as a function of top or bottom FeB insertion thickness under different annealing conditions. For top FeB insertion (figure 5(a)), the influence of FeB insertion on $\alpha$ shows a moderate dependence on its thickness after 350 °C annealing. For bottom FeB insertion (figure 5(b)), however, $\alpha$ reaches the minimum at 0.4 nm FeB insertion and increases dramatically beyond the measurement range with the FeB thickness in both annealing conditions. For those cases, the resonance is too broadened to be resolved, reflecting a very large damping [21]. It also leads to a huge uncertainty in the resonance field and hence $\mu_0 H_k^{eff}$ determination, as mentioned in the preivous section.

An increase of $\alpha$ is observed at 0.6 nm FeB in the 400 °C annealing condition. However, a reduction in $\alpha$ after 400 °C annealing is obtained when ultrathin FeB (0.2 – 0.4 nm) is inserted at either interface. This suggests that the annealing treatment has different effects on $\alpha$ of the samples with various FeB insertion thicknesses. Perhaps different amount of FeB insertion leads to changes in the Fe concentration, microstructures and crystallization of dual-MgO FLs after boron depletion upon annealing and thereby to different damping behaviors [29,30].

Finally, Table 1 summarizes and compares MTJ stacks after 350 °C annealing with FeB insertion at top, bottom, or both CoFeB/MgO interfaces. As the FeB insertion thickness and position differ, the PMA of the dual-MgO FL can be tuned in a wide range, while the damping is almost independent. From the systematic studies on insertion thickness in the previous sections, top and bottom FeB are optimized to be 0.2 nm and 0.4 nm, respectively. As a result, the MTJ stack with such FeB insertion at both interfaces in dual-MgO FLs can be engineered with a high TMR, low RA, large $K_U^{eff} \cdot t$, low $M_s \cdot t$, high $\mu_0 H_k^{eff}$, and low damping constant.

## 4. Conclusion

In this paper, we explore the impact of Fe$_{80}$B$_{20}$ layer inserted at two interfaces of Co$_{40}$Fe$_{60}$B$_{20}$/MgO in dual-MgO FLs in MTJ stacks and its annealing stability. With ultrthin FeB (0.2 – 0.4 nm) inserted at the top or bottom CoFeB/MgO interface, the TMR can be maintained with lower RA values, while the top-FeB insertion results in a more RA drop with a similar TMR. In both cases, the FL saturation magnetization reduces with increasing the inserted FeB thickness, while the FL effective anisotropy field increases. However, the PMA of dual-MgO FLs with FeB inserted at the bottom interface shows a larger improvement than its top FeB insertion counterpart, even after 400 °C annealing. At the same time, the FeB (0.2 − 0.4 nm) insertion at either interface reduces the damping constant in the FL. By optimizing the FeB insertion layer thickness, the dual-MgO FL with a low saturation magnetization, high effective anisotropy field and low damping can be achieved after 400 °C annealing. However, the performance degrades if a thicker FeB is used to replace CoFeB in dual-MgO FLs.

This study demonstrates a novel approach to tune dual-MgO FL properties other than typical boron composition or non-magnetic spacer engineering. By using the FeB insertion layer at CoFeB/MgO interfaces, magnetic properties of the FL and the magnetotransportation of MTJs can be engineered in a wide range, which enables MTJs to meet different performance requirements for various spintronic applications.

**Acknowledgements**

This work was supported by NRF Investigatorship (NRFI06-2020-0015).

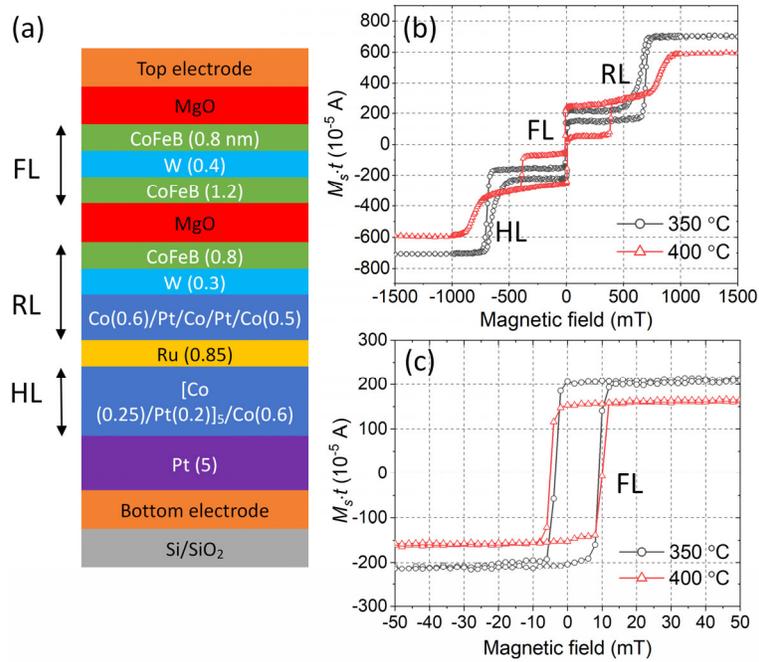

**Figure 1.** (a) Stack layout of blanket MTJs without FeB insertion. Free layer (FL), reference layer (RL) and hard layer (HL) are indicated. Thicknesses of sublayers are shown in nm with parentheses. Blanket films were annealed first at 350 °C and then at 400 °C, both for 30 min. (b) Major loop and (c) minor loop of the stack in (a) measured by VSM after different annealing conditions with the magnetic field perpendicular to the sample plane.

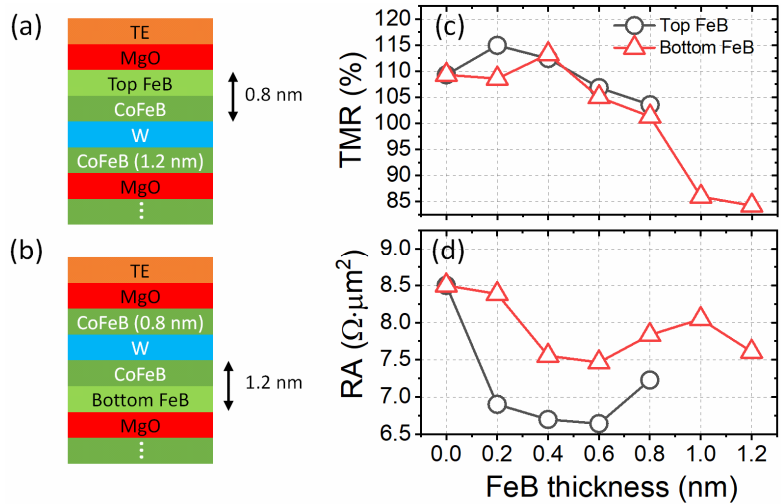

**Figure 2.** Schematic of FeB insertion at (a) top and (b) bottom CoFeB/MgO interface in dual-MgO FL in the stack shown in figure 1(a). The total thickness of inserted FeB plus remaining CoFeB is kept at 0.8 nm and 1.2 nm for layers above and below W spacer, respectively. The impact of FeB insertion on TMR (c) and RA (d) of the MTJ stacks after 350 °C annealing are shown.

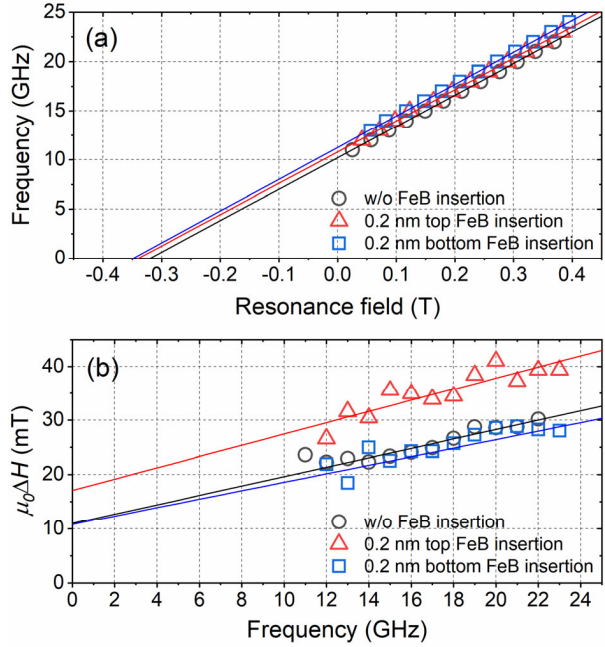

**Figure 3.** (a) The external excitation frequency as a function of ferromagnetic resonance field and (b) the linewidth versus frequency of FL in MTJ stacks without FeB insertion (black open circles), with 0.2 nm top FeB insertion (red open triangles), and with 0.2 nm bottom FeB insertion (blue open squares). Solid lines are fits.

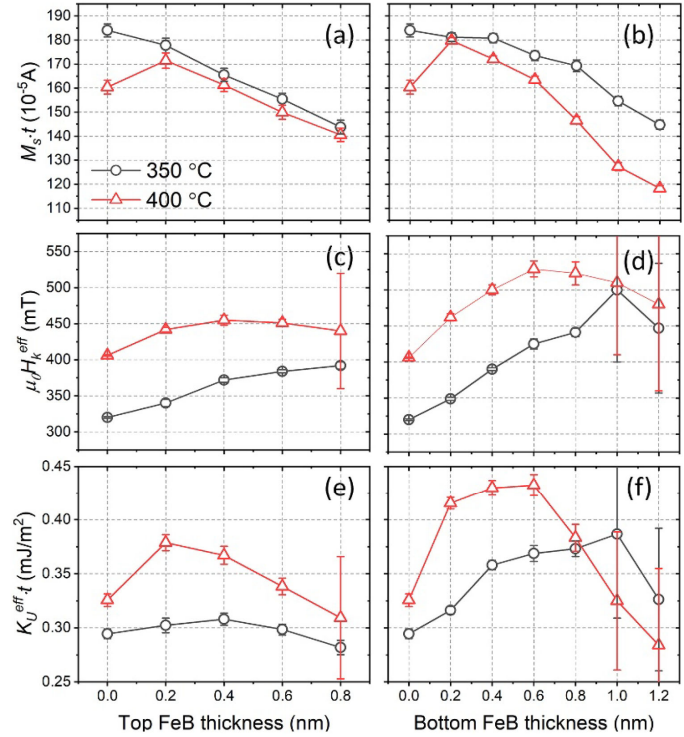

**Figure 4.** Effect of FeB insertion and annealing conditions on $M_s \cdot t$ ((a) and (b)), $\mu_0 H_k^{eff}$ ((c) and (d)), and $K_U^{eff} \cdot t$ ((e) and (f)). (a), (c) and (e) show the impact from FeB insertion at the top CoFeB/MgO interface, while (b), (d) and (f) show the impact from bottom interface.

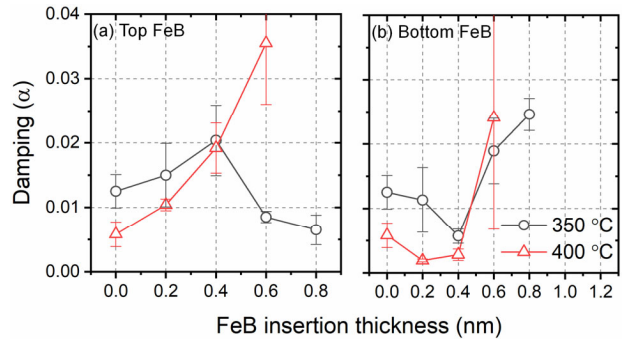

**Figure 5.** Gilbert damping as a function of FeB insertion thickness at (a) top interface and (b) bottom interface under two annealing conditions.